\documentclass[12pt]{article}
\usepackage{amssymb,amsmath,graphicx,natbib,titlesec,url,enumitem,color}
\usepackage[colorlinks=true]{hyperref}
\hypersetup{
    citecolor={blue},
    urlcolor={blue}
    }
\usepackage{times}
\usepackage{geometry}
\usepackage[utf8]{inputenc}
\usepackage{enumitem,amssymb}
\usepackage{ragged2e}
\newlist{thematic}{itemize}{8}
\setlist[thematic]{label=$\square$}
\usepackage{pifont}

\newcommand{\arcsec}{\mbox{$^{\prime\prime}$}}
\newcommand{\apj}{ApJ}

\newcommand{\aap}{A\&A}
\newcommand{\mnras}{MNRAS}
\newcommand{\apjs}{ApJSS}
\newcommand{\pasp}{PASP}
\newcommand{\araa}{ARAA}
\newcommand{\nat}{Nature}
\newcommand{\apjl}{ApJ}

\titleformat{\section}[block]
{\normalfont\filcenter\sffamily}{}{.5em}{\bfseries}

\titleformat{\subsection}[block]
{\normalfont\sffamily}{}{}{\bfseries}

\begin{document}
\raggedright
\huge
Astro2020 Science White Paper \linebreak

Cold Gas Outflows, Feedback, and the Shaping of Galaxies \linebreak
\normalsize

\noindent \textbf{Thematic Areas:} \hspace*{60pt} $\square$ Planetary Systems \hspace*{10pt} $\square$ Star and Planet Formation \hspace*{20pt}\linebreak
$\square$ Formation and Evolution of Compact Objects \hspace*{31pt} $\square$ Cosmology and Fundamental Physics \linebreak
  $\square$  Stars and Stellar Evolution \hspace*{1pt} $\square$ Resolved Stellar Populations and their Environments \hspace*{40pt} \linebreak
  $\checkmark$    Galaxy Evolution   \hspace*{45pt} $\square$             Multi-Messenger Astronomy and Astrophysics \hspace*{65pt} \linebreak
  
\textbf{Principal Author:}

Name: Alberto D. Bolatto	
 \linebreak						
Institution: University of Maryland, Department of Astronomy, College Park, MD 20742, USA 
 \linebreak
Email: bolatto@astro.umd.edu
 \linebreak
Phone: +1-301-405-1521 
 \linebreak
 
\textbf{Co-authors:} 
Lee Armus${}^2$, Sylvain Veilleux${}^1$, Adam K. Leroy${}^3$, Fabian Walter${}^4$, Richard Mushotzky${}^1$, Karin M. Sandstrom${}^5$, Paul Martini${}^{3}$, Evan E. Schneider${}^6$, Tony Wong${}^7$, Roberto Decarli${}^8$, Caitlin Casey${}^9$, Dominik Riechers${}^{10}$, David Meier${}^{11}$, Desika Narayanan${}^{12}$

\bigskip
$^1$University of Maryland, Department of Astronomy, College Park, MD 20742, USA; $^2$Infrared Processing and Analysis Center, California Institute of Technology, Pasadena, CA 91125, USA; $^3$Ohio State University, Department of Astronomy, Columbus, OH 43210, USA; $^4$Max Planck Institute for Astronomy, Heidelberg, Germany; $^5$University of California at San Diego, Department of Physics, La Jolla, CA 92093; $^6$Princeton University, Department of Astrophysical Sciences, Princeton, NJ 08544; $^7$University of Illinois, Department of Astronomy, Urbana, IL 61801; $^8$Italian National Institute for Astrophysics, Astrophysics and Space Science Observatory, Bologna, Italy; $^9$The University of Texas, Department of Astronomy, Austin, TX 78712; $^{10}$Cornell University, Department of Astronomy, Ithaca, NY 14853; $^{11}$New Mexico Tech, Department of Physics, Socorro, NM 87801; $^{12}$University of Florida, Department of Astronomy, Gainesville, FL 32611

\bigskip
\textbf{Abstract:}
There is wide consensus that galaxy outflows are one of the most important processes determining the evolution of galaxies through cosmic time, for example playing a key role in shaping the galaxy mass function. Our understanding of outflows and their drivers, however, is in its infancy --- this is particularly true for the cold (neutral atomic and molecular) phases of outflows, which present observational and modeling challenges. Here we outline several key open questions, briefly discussing the requirements of the observations necessary to make progress, and the relevance of several existing and planned facilities. It is clear that galaxy outflows, and particularly cold outflows, will remain a topic of active research for the next decade and beyond. 

\bigskip
\textbf{Related white papers:} See also white papers led by M. Ruszkowski and K. Nyland.   
\pagebreak




\justify



It has become increasingly clear that it is impossible to understand how galaxies form and evolve through cosmic time without a deeper understanding of how 
feedback from star formation and black hole accretion affect the growth of galaxies and the state of their gas reservoirs. While much progress has been made on both large-scale cosmological simulations and detailed simulations of individual galaxies, the continued struggle to reproduce the observed characteristics of the galaxy population establishes feedback as 
the key open question in galaxy evolution. This is intimately linked to one of the fundamental questions identified in the {\em ``New Worlds, New Horizons''} decadal report: how do baryons cycle in
and out of galaxies? 

Galactic winds are thought to shape the galaxy mass function, heat the circumgalactic medium (CGM), play a critical role in quenching star formation --- through gas removal, heating, and moment deposition ---, and pollute the intergalactic medium  \citep[e.g.,][]{VEILLEUX2005}.
Galaxies are not closed systems, and winds are necessary to explain chemical evolution, as well as playing a key role in shaping the disk-halo interface. The fastest, most energetic winds arise due to active galactic nuclei (AGN) or strong starbursts, while less powerful, more localized galactic fountains due to clustered star formation recycle material between the disk and the halo. 
Winds do not occur in isolation. Directly and through the wind, the same engines that power winds exert very significant feedback on the cold ISM of the galaxy --- the material for further star formation --- through the input of energy and momentum. 
It is their complexity and importance that makes winds, and their engines, fertile areas for study over the next decade and beyond.

Detailed studies of nearby systems frequently focused on the warm or hot phases of winds, which are often visible in X-rays or optical emission lines. 
The manifestation of these hot winds can often be quite spectacular, consisting of huge, bipolar nebulae \citep[e.g.,][]{HECKMAN1990,STRICKLAND2004}, but they appear to carry little mass $M_{Xray}\sim10^6$\,M$_\odot$ \citep[e.g.,][]{LEHNERT1999}. 
Galactic winds, however, are multi-phase phenomena and the hot/warm phases are only part of the story.
When present, colder phases, constituted by denser neutral atomic and molecular gas, can dominate the mass and metal budget of the outflow \citep{WALTER2002,RUPKE2005,FERUGLIO2010,ALATALO2011,RUPKE2013,RUPKE2017}. 

Until recently, observations of the cooler phases of galactic outflows have been hindered by a lack of sensitivity and/or spatial resolution to properly image the low surface brightness wind and unambiguously connect it to the processes in the disk that power the outflow.  With {\em Herschel}, it was possible to detect and model fast, dense outflows in a number of local Ultraluminous Infrared and starburst galaxies \citep[e.g.,][]{STURM2011,VEILLEUX2013,GONZALEZ-ALFONSO2017}.  The derived mass outflow rates are, in some cases, comparable to or larger than the star formation rates of the galaxies themselves, implying a significant impact on the lifetime of the active phase. However, {\em Herschel} could only detect the nearest, brightest sources, and the winds were nearly always unresolved. Radio interferometers fare better at resolving cold winds. ALMA and NOEMA are currently making important advances in finding and imaging molecular outflows \citep{BOLATTO2013a,COMBES2013,CICONE2014,SAKAMOTO2014,GARCIA-BURILLO2014,ZSCHAECHNER2016,VEILLEUX2017,GOWARDHAN2018,FLUETSCH2019}, but because of the resolution and surface brightness sensitivity requirements, identifying, characterizing, and obtaining good statistics on molecular winds remain very challenging with present-day facilities.

\section{Open Questions on Cold Galaxy Outflows}

Despite the progress in studying and modeling cold winds in the last several years, a number of key outstanding questions remain. Because of their significance to galaxy evolution, these questions will be the likely focus of much research during the next decade.

\vspace{0.1cm}
\noindent{\bf What are the wind mass loss rates and efficiencies?}\\ 
What is the mass-loss to star-formation-rate ratio (the mass loading parameter)? Are mass-loading parameters and mass-loss rates in low-mass galaxies as high as predicted? These parameters are key inputs to cosmological simulations, necessary to understand the precise effects of feedback on galaxy growth. Current estimates of the efficiency with which momentum is imparted to the different gas phases, implemented as sub-grid recipes in physical galaxy simulations, suggest mass loading parameters of order $\eta\sim3-10$ for massive galaxies, and as high as $\eta\sim100$ for dwarf galaxies \citep{MURATOV2015}. While higher mass galaxies are expected to retain most of their metals in their circum-galactic environment, dwarf galaxies should {heavily pollute} the IGM \citep{MURATOV2017}. Such values of $\eta$ appear to be necessary to reproduce galaxy properties in cosmological simulations. It remains unclear, however, how to precisely attain such high mass loading efficiencies in detailed simulations \citep{KIM2018}. It is likely that most of the mass loss is due to the denser, cold phases, likely dominated by the neutral atomic and molecular gas and thus directly observable at mid- to far-infrared, millimeter-wave, and radio bands. Some of the key information needed to measure accurate molecular mass loss rates is an understanding of the column density and volume density structure in the wind, critically dependent on observations of dust continuum and molecular species in the outflowing gas. Other uncertainties are associated with velocities and geometry, which can be constrained through models \citep[e.g.,][]{STURM2011}, sometimes using optical observations \citep[e.g.,][]{WESTMOQUETTE2011}.

\vspace{0.1cm}
\noindent{\bf What fraction of the outflowing gas escapes the galaxy depending on galaxy properties?}\\ 
Line-of-sight absorption spectroscopy of starburst galaxies finds that the warm/hot outflow velocities are typically comparable to the escape velocities for $L^*$ galaxies, and substantially exceed them for dwarfs \citep{HECKMAN2000}. This dependence on galaxy properties has important implications for the metal enrichment of the circumgalactic and intergalactic medium \citep{WERK2016,TUMLINSON2017} and the mass-metallicity relationship \citep{PEEPLES2011}. 
Studies of cool outflows, however, are more ambiguous about the fraction of mass that escapes galaxies \citep{LEROY2015,WALTER2017,MARTINI2018}.
The question of escape fraction could be better answered with more observations that trace the wind kinematics well outside the star forming disk, combined with better models for the halo mass distribution of the starbursts.

\vspace{0.1cm}
\noindent{\bf What is the role of any reaccreted/recycled gas?}\\ 
Any wind material that falls back onto the galaxy provides new fuel for star formation, and recycled material may play an important role in lengthening the gas depletion timescale of galaxies \citep{DAVE2011}, feeding galaxies at late cosmic times, and {allowing for the exchange of} processed gas between galaxies \citep{ANGLES2017b}. Detailed simulations suggest that winds and fountains go through phases, dominated alternatively by outflow and inflow \citep{KIM2018}. Generally, understanding the fraction of escaping gas and the relative amounts of expelled and recycled material requires high sensitivity observations of gas in galaxies and their circum-galactic environments. More sensitive, large-scale radio observations could also detect any cold, fountain material raining back down on galaxy disks. 

\vspace{0.1cm}
\noindent{\bf What are the statistical properties of winds in the universe?} \\
Are these a rare phenomenon confined to AGN and starbursts, or are they a general feature of galaxies? What is the redshift evolution of starburst- and quasar-driven winds? Observations suggest that galactic winds are ubiquitous at high redshift \citep[e.g.,][]{NEWMAN2012}. We know very little about their cold components, except in a handful of spectacular examples \citep[e.g.,][]{MAIOLINO2012}. Fast outflows are seen in powerful IR galaxies in the local Universe \citep{VEILLEUX2013,PEREIRA-SANTAELLA2018}, but studies have been limited to a handful of the brightest sources. Large, sensitive, systematic mm-wave, far-infrared, and X-ray spectroscopic surveys reaching out to epochs where galaxies are rapidly evolving are necessary to determine links between outflow and galaxy properties, and establish their importance in a cosmological context.  

\vspace{0.1cm}
\noindent{\bf What are the driving mechanisms of cold winds?}\\
Mechanical feedback from supernovae \citep[e.g.,][]{FUJITA2009}, radiation pressure \citep[e.g.,][]{MURRAY2011}, cosmic ray streaming \citep[e.g.,][]{RUSZKOWSKI2017}, entrainment facilitated by Kelvin-Helmholtz instabilities \citep[e.g.,][]{HECKMAN2000}, and direct driving by interaction with AGN jets \citep[e.g.,][]{WAGNER2011} have all been proposed as ways to inject momentum in the gas. It remains unclear, however, how they combine, which one if any dominates, and how that evolves through cosmic time \citep{HOPKINS2012,MURATOV2015,ROSDAHL2015}. A combination of high-resolution and sensitivity observations of the cold phases as they are ejected are key to solve this problem. For example, it appears that radiation pressure is insufficient to explain the high-resolution properties of the starburst-driven NGC~253 molecular outflow \citep{WALTER2017}. High resolution imaging of the cold gas clumps can provide information about the type of forces they are experiencing, particularly in comparison to simulations. Velocity measurements can uncover accelerations, which will be also a clue for the launching processes.  

\vspace{0.1cm}
\noindent{\bf What are the conditions triggering cool outflows?}\\
Observations suggest the existence of a star formation surface density threshold for launching large outflows \citep[e.g.,][]{NEWMAN2012}, and a similar threshold in luminosity appears to exist for AGN-launched outflows \citep[e.g.,][]{VEILLEUX2013,WYLEZALEK2017}. Lower velocity galactic fountains and even radiation pressure-driven outflows, can also occur over extended areas of disks. Systematic demographics of the different properties of the multiphase outflows and their hosts, collected through a combination of multi-wavelength observations will provide valuable information about the conditions and triggers of outflows.

\vspace{0.1cm}
\noindent{\bf What is the distribution of the phases of the wind? }\\
Imparting momentum to molecular cloudlets without destroying them has proven difficult in simulations \citep{SCANNAPIECO2015,BRUGGEN2016}. Cold gas destruction may mass-load the hot phase \citep{SCHNEIDER2017}, while at the same time part of the hot phase outflow may cool and reform a cold phase \citep[e.g.,][]{THOMPSON2016,SCHNEIDER2018}. Observations of the outflow in M~82 strongly suggest conversion of molecular into neutral atomic gas as the outflowing gas progresses away from the galaxy \citep{LEROY2015}. But molecular outflows also show evidence for rich chemistry \citep[e.g.,][]{LINDBERG2016,WALTER2017}. What is the temperature and shock structure in the outflowing gas? Is it heated as part of the ejection process \citep{DASYRA2014}? High sensitivity, resolved imaging in neutral atomic gas, dust, and molecular tracers is needed to answer these questions.


\vspace{0.1cm}
\noindent{\bf What feedback effects do winds exert on the host galaxy ISM?}\\
Winds not only eject material from their host galaxy, but in doing so they inject energy (in the form of turbulence) and momentum in the neighboring material. 
The overpressured hot cavity caused by the engine is surrounded by a compressed, dense shell of material likely stripped by Kelvin-Helmholtz instabilities and related thermal processes associated to the shear with the hot flow \citep{HECKMAN2000}. In regions shocked by the winds, temperature and turbulence are elevated, and star formation activity per unit gas appears to drop accordingly \citep{GUILLARD2012,GUILLARD2015}. 
The relative importance of this feedback compared to the other effects of outflows (direct mass loss, suppression of further accretion, lengthening time-scales due to recycling) remains to be understood and characterized. Resolved studies of nearby galaxies at radio, mm-wave, and far- to mid-infrared wavelengths are needed to improve our understanding of these processes. 

\vspace{0.1cm}
\noindent{\bf Are winds only effective at suppressing star formation, or can they also trigger it?}\\
Although outflows driven by star formation or AGN are frequently invoked as mechanisms to expel gas and/or quench star formation \citep[e.g.,][]{ALATALO2015}, star formation can also be enhanced by compressive turbulence driven by the mechanical energy input of the wind \citep[e.g.,][]{VANBREUGEL1985,CROFT2006}. These positive feedback processes can also have an important effect on galaxy evolution \citep{SILK2013}. Multi-wavelength radio and far-infrared techniques are particularly well suited to studying this problem, since they can measure extincted star formation rates as well as image the outflowing gas and measure its kinematics.

\vspace{0.1cm}
\noindent{\bf How do winds affect the growth of black holes?}\\
Simulations suggest that black hole growth, particularly at early times, is limited by stellar feedback, which expels gas from galactic nuclei, limiting accretion. As a consequence, black holes can be under-massive in low-mass galaxies with respect to their high-mass counterparts, causing them to fall below the M$_{BH}-\sigma_{halo}$ relation \citep[e.g.,][]{ANGLES2017a}. Addressing these questions requires multi-wavelength high angular resolution observations of the highly extincted central regions of galaxies, including X-rays, mid- to far-infrared, and radio data.

\vspace{0.1cm}
\noindent{\bf What is the relation between the large scale fast-moving cold outflows and their AGN?}\\ 
Are the ultra-fast outflows (UFOs) seen in the X-rays on sub-parsec scale the ultimate drivers of the most powerful cold outflows, or are the more common but slower soft X-ray and ultraviolet (UV) warm absorbers and UV broad absorption line (BAL) outflows (seen on larger scale than the UFOs) a better predictor of these cold outflows? How is the energy in these nuclear winds transferred to drive the galaxy-scale cold outflows? 
In theory, very fast accretion-disk winds indentified in the X-rays \citep{REEVES2003,TOMBESI2010} can drive shocks into the host galaxy ISM and create shock-driven over-pressurized bubbles that give rise to the large-scale outflows observed in ionized, neutral, and molecular gas \citep{FAUCHER-GIGUERE2012,TOMBESI2015}.  
Directly linking fast accretion disk outflows with galactic winds  requires high-quality, velocity-resolved imaging of the molecular and neutral atomic gas at high spatial resolution in the central regions of AGN with identified X-ray UFOs. A more complete survey of AGN will help relate the cold outflows to the more prevalent warm absorbers \citep{CRENSHAW2012} and BAL outflows  \citep{GIBSON2009}. The prospects are good that radio and future UV-optical facilities will be able to spatially resolve the regions where the BAL outflows interact with the ISM of the AGN host galaxies \citep{MOE2009,BAUTISTA2010,DUNN2010}.

\section{Prospects for Moving Forward in the Next Decade and Beyond}

Understanding galactic outflows requires a combined, coordinated effort to further the physical modeling of the phenomenon with a host of new observations necessary to provide the boundary conditions to the problem. 
To answer the open questions posed above we need both large area sensitive spectroscopic surveys at mid- to far-infrared wavelengths and high resolution radio-mm and X-ray observations. 
At radio and millimeter wavelengths this implies using interferometers from the ground. At the shorter mid- to far-infrared wavelengths the closest examples can be studied at good spatial resolution with single-aperture space telescopes, which when equipped with efficient detectors are also capable of carrying out large surveys for more distant sources. 

Existing facilities, such as ALMA which will operate over the next decade and beyond, will continue to provide invaluable observations. ALMA enables sensitive high-resolution imaging, particularly at wavelengths of 1~mm and shorter, which provide access to several molecules including the J=$2-1$ and $3-2$ transitions of CO --- the main tracer of molecular material. ALMA has produced a future development roadmap\footnote{\url{https://www.almaobservatory.org/en/publications/the-alma-development-roadmap}}. Among the recommended enhancements, the wide-banding of existing receivers will speed up multi-transition cold outflow observations. 

Among existing concepts for future facilities two of them, the {\em Next Generation Very Large Array} (ngVLA\footnote{\url{http://ngvla.nrao.edu}}) and the {\em Origins Space Telescope}\footnote{\url{https://origins.ipac.caltech.edu}}, stand out as likely to produce the most interesting observational breakthroughs for the cold winds themselves and the starburst engines, while a third concept, the {\em Lynx X-ray Observatory}\footnote{\url{https://www.lynxobservatory.com}}, is particularly promising in terms of delivering ground-breaking observations for the central AGN engines and the hot phases of the outflows.

The ngVLA concept will provide an order-of-magnitude improvement in collecting area over the VLA. Operating between 1 and 116 GHz, it will also yield roughly an order-of-magnitude sensitivity improvement over ALMA in the overlapping range of frequencies. Crucially, the ngVLA will deliver 21~cm atomic hydrogen, 18~cm OH and 9~cm CH (faint but good tracers of low density molecular gas), and 2.6~mm CO J=$1-0$ molecular imaging that is well beyond the capabilities of any instrument existing or in construction. It will also provide access to a large number of spectroscopic diagnostics for the outflowing molecular and atomic gas (density, shocks, column density, abundance, and temperature tracers), the faintness of which currently precludes observation (for example, SiO, a tracer produced by dust destruction in strong shocks). 
The phase-1 of the Square Kilometer Array (SKA) currently in construction will be able to access the atomic hydrogen and OH transitions in southern wind-hosting galaxies, albeit with lower sensitivities and in general more limited capabilities than anticipated for the ngVLA. However, its operation is limited to the lower frequencies and it will never be able to observe CO and most molecular tracers, a limitation that will be also shared by the full phase-2 SKA if it is ever built. Note also that as the US is not a partner, the access of US investigators remains unclear and it will likely be limited.

The {\em Origins} concept will provide a huge leap in spectroscopic and continuum sensitivity over previous far-infrared missions, with better mapping speed and angular resolution. {\em Origins} enables access to mid- and far-infrared tracers of feedback and heating by young stars and AGN,
as well as OH and H$_2$O molecular lines that trace cool outflows, at the sensitivity required to observe them for extragalactic and high-$z$ sources out to $z\sim4-5$. The combination of fast-mapping and sensitivity enables the large surveys needed to characterize galaxy populations, as well as quality spectroscopic imaging of nearby targets. The expected sensitivity for the [CII] 158 $\mu$m transition, for example, will make it an unparalleled instrument for imaging low surface brightness extended material around galaxies, resulting from fountain and outflow activity. The proposed JAXA/ESA mission SPICA, with less sensitivity, slower mapping speed, and a shorter wavelength cutoff than {\em Origins}, would constitute a more limited but still capable observatory for studying outflows.

The {\em Lynx} concept offers very high angular resolution ($0.5\arcsec$) with very high spectral resolution that enables it to characterize the composition, ionization state, and --- very importantly --- the kinematics of the hot phases of the winds. The planned sensitivity of Lynx will allow it to study M~82 like starbursts at moderate redshifts, and the most luminous of starbursts will be detectable out to $z\sim2-3$. The high sensitivity and resolution in the Fe K band will enable detailed analysis of the central AGN engines and associated UFOs, including their energetics and geometry. The ESA mission ATHENA, with much lower spatial and spectral resolution than Lynx, is suited to studying outflows from powerful AGN but will not be able to do so for slower outflows (which includes all starburst-driven outflows).




\pagebreak

\end{document}